\documentclass[aps,pra,twocolumn,twoside,preprintnumbers,superscriptaddress,
               nofootinbib]{revtex4}

\usepackage{amsthm,amssymb}

\renewcommand{\>}{\rangle}

\newcommand{\ket}[1]{\mbox{$| #1 \rangle$}}

\newcommand{\be}{\begin{equation}}
\newcommand{\ee}{\end{equation}}
\newcommand{\bea}{\begin{eqnarray}}
\newcommand{\eea}{\end{eqnarray}}
\newcommand{\nn}{\nonumber \\}

\renewcommand{\H}{{\cal H}}
\newcommand{\R}{\mathbb{R}}

\newcommand{\g}{\raisebox{0.3ex}{$\gamma$}}
\newcommand{\s}[1]{\sigma_{#1}}

\newcommand{\tr}{\mathop{\mathrm{tr}}\nolimits}
\newcommand{\diag}{\mathop{\mathrm{diag}}\nolimits}

\newcommand{\bigboxplus}{\mathop{\mbox{\large $\boxplus$}}\displaylimits}

\newtheorem{theorem}{Theorem}
\newtheorem{lemma}[theorem]{Lemma}
\newtheorem{corollary}[theorem]{Corollary}
\newtheorem{arule}{Rule}
\newtheorem{prop}{Property}
\newtheorem*{axiom}{Axiom}

\begin{document}

\title{Reversible simulation of bipartite product Hamiltonians}

\author{Andrew M. Childs}
\email[]{amchilds@mit.edu}
\affiliation{Center for Theoretical Physics,
             Massachusetts Institute of Technology,
             Cambridge, MA 02139, USA}

\author{Debbie W. Leung}
\email[]{wcleung@cs.caltech.edu}
\affiliation{Institute for Quantum Information,
             California Institute of Technology,
             Pasadena, CA 91125, USA}

\author{Guifr\'e Vidal}
\email[]{vidal@cs.caltech.edu}
\affiliation{Institute for Quantum Information,
             California Institute of Technology,
             Pasadena, CA 91125, USA}

\date[]{14 March 2003}

\preprint{MIT-CTP \#3347}


\begin{abstract}
Consider two quantum systems $A$ and $B$ interacting according to a
product Hamiltonian $H=H_{A}\otimes H_{B}$.  We show that any two such
Hamiltonians can be used to simulate each other {\em reversibly}
(i.e., without efficiency losses) with the help of local unitary
operations and local ancillas. 
Accordingly, all non-local features of a product
Hamiltonian---including the rate at which it can be used to produce
entanglement, transmit classical or quantum information, or simulate
other Hamiltonians---depend only upon a single parameter. We identify
this parameter and use it to obtain an explicit expression for the
entanglement capacity of all product Hamiltonians.  Finally, we show
how the notion of simulation leads to a natural formulation of
measures of the strength of a nonlocal Hamiltonian.
\end{abstract}

\maketitle

\section{Introduction}
\label{sec:intro}

Suppose two quantum systems $A$ and $B$ are coupled by some nontrivial
interaction Hamiltonian $H \neq H_A \otimes I_B + I_A \otimes H_B$.
Such a Hamiltonian can be used for a variety of information processing
tasks, such as transmitting classical or quantum information, creating
entanglement, or simulating other nonlocal evolutions.  One of the
goals of quantum information theory is to quantify the capacity of an
interaction to perform such information processing tasks. 

The various nonlocal properties of Hamiltonians can be analyzed in
different ways, but they are typically studied under a common
framework of perfect local control.  To focus on the uniquely nonlocal
features of Hamiltonians, all local abilities are regarded as free
resources.  This includes the use of local ancillary degrees of
freedom and the ability to perform fast local operations to modify the
evolution.

One of the tasks mentioned above consists of using $H$ and local
operations to simulate another Hamiltonian $H'$, a control technique
that allows one to modify a naturally available interaction into a
more convenient one.
Any bipartite interaction Hamiltonian can simulate any other at some
nonzero rate \cite{DNBT02,BCL02,WRJB02,NBD02}.
However, we would ultimately like to know the {\em most efficient} way
to use $H$ to simulate $H'$, as well as the optimal rate $\g_{H'|H}$
at which the simulation can be accomplished.
A method for optimal simulation of two-qubit Hamiltonians is given in
\cite{BCL02}, and the optimal rate in this case can be expressed in
terms of a majorization condition \cite{VC02}.  However, to the best
of our knowledge, no optimal simulation rates have been reported
beyond the two-qubit case.

Understanding Hamiltonian simulation also provides insight into
capacities for other information processing tasks.  Let $C_H$ denote
the capacity of Hamiltonian $H$ to accomplish one of these tasks,
again assuming perfect local control.
If Hamiltonian $H$ can be used to simulate $H'$ at a rate $\g_{H'|H}$,
then
\be
  C_{H} \ge \g_{H'|H} \; C_{H'} 
\,,
\label{eq:simbound}
\ee
since one could first use $H$ to simulate $H'$ and then use $H'$ to
accomplish the task.  Equation (\ref{eq:simbound}) is a lower bound on
the capacity of $H$, or equivalently, an upper bound on the capacity
of $H'$. Of course, such bounds need not be tight.  For example, the
majorization condition for optimal simulation of two-qubit
Hamiltonians only provides a partial order on these Hamiltonians, and
thus the resulting bounds on capacities---for example, on the
entanglement capacity \cite{DVCLP01,BHLS02,CLVV03}---are not always
tight.

The fact that any nonlocal Hamiltonian can simulate any other at some
nonzero rate means that all interactions are {\em qualitatively}
equivalent.
A much stronger, {\em quantitative} notion of equivalence between
interactions comes from the possibility of performing a {\em
reversible} simulation.  We say that $H$ and $H'$ can simulate each
other reversibly if we can use $H$ to simulate $H'$, and then use $H'$
to simulate $H$ back, with no overall loss in efficiency. In terms of
simulation rates, reversible simulation amounts to the condition
\be
  \g_{H|H'} \, \g_{H'|H} = 1
\,.
\label{eq:reversible}
\ee
Notice that if two Hamiltonians $H$ and $H'$ can simulate each other
reversibly, then their capacities are related by
\be
  C_H = \g_{H'|H} \, C_{H'}
\,,
\label{eq:rev}
\ee 
as can be seen by applying (\ref{eq:simbound}) in both directions.
Furthermore, if every pair of Hamiltonians in some given set can
simulate each other reversibly, then simulation provides a total order
on the set.  Thus the nonlocal properties of the entire set can be
studied by focusing on only one Hamiltonian in the set. 

In this paper we consider the set of bipartite Hamiltonians that can
be written as a tensor product of the form
\be
  H = H_A \otimes H_B
\,,
\label{eq:product}
\ee
where $H_A$ acts on system $A$ and $H_B$ acts on system $B$. We shall
call such a Hamiltonian a {\em product Hamiltonian} for short.  An
example of a product Hamiltonian in a two-qubit system is the Ising
interaction $H_{\rm Ising}$,
\be
  H_{\rm Ising} := \s{z} \otimes \s{z}, \qquad
          \s{z} := \left[ \matrix{1 & 0 \cr 0 & -1} \right]
\,.
\label{eq:ising}
\ee 
Our main result is an explicit protocol for the reversible simulation 
of any product Hamiltonian by another. 
It follows that the nonlocal properties of a product Hamiltonian $H$
depend entirely on a single parameter.  We denote this parameter by
$K_{\otimes}(H)$, and choose it to be the rate $\g_{H_{\rm Ising}|H}$
at which $H$ can simulate the Ising interaction.  We find that
\be
  K_{\otimes}(H) = {1 \over 4} \Delta_A \Delta_B
\,,
\label{eq:Ising}
\ee
where $\Delta_{A}$ ($\Delta_B$) denotes the difference between the
largest and the smallest eigenvalues of $H_{A}$ ($H_B$). The optimal
simulation rate between any two product Hamiltonians $H$ and $H'$ can
be written in terms of $K_{\otimes}$ as 
\be
  \g_{H'|H} = \frac{K_{\otimes}(H)}{K_{\otimes}(H')}
\,,
\label{eq:gHH}
\ee 
so that any capacity $C_H$ known for just one product Hamiltonian can
be easily computed for any other product Hamiltonian using
(\ref{eq:rev}) and (\ref{eq:gHH}).
In particular, we use previous results for the Ising interaction
\cite{CLVV03} to obtain a simple expression for the entanglement
capacity of any product Hamiltonian.

In addition to quantifying the ability of an interaction to perform
particular tasks, one can imagine defining abstract measures of the
nonlocality of an interaction.  Reference \cite{NDD02} introduced the
notion of a {\em strength measure} $K(U)$ of a nonlocal quantum
operation.  Three axioms that every strength measure should satisfy
were proposed, along with several additional desirable properties.
Here we consider strength measures $K(H)$ for nonlocal Hamiltonians.
We will formulate a single axiom in terms of Hamiltonian simulation
that implies many desirable properties.
For the special case of product Hamiltonians, our results imply an
essentially unique measure, $K_{\otimes}(H)$ in (\ref{eq:Ising}), with
several additional properties.

The structure of the paper is as follows.  In Section \ref{sec:sim},
we describe the problem of bipartite Hamiltonian simulation in more
detail and discuss a set of basic simulation rules that can be
used to build up all possible simulations by composition.  In Section
\ref{sec:equiv}, we derive our main result, namely that all tensor
product Hamiltonians can reversibly simulate each other.  In Section
\ref{sec:ent}, we apply this result to the calculation of entanglement
capacities, and in Section \ref{sec:strength}, we relate Hamiltonian
simulation to the strength measure formalism.  Finally, in Section
\ref{sec:discussion}, we conclude with a discussion of our findings
and of some open problems.

\section{Simulating bipartite Hamiltonians}
\label{sec:sim}

The problem of bipartite Hamiltonian simulation can be posed as
follows.  We consider two quantum systems $A$ and $B$ held by Alice
and Bob, respectively. The systems interact according to some nonlocal
Hamiltonian $H\neq H_A \otimes I_B + I_A \otimes H_B$.  Alice and Bob
want to use $H$ to produce an evolution according to some other
bipartite Hamiltonian $H'$.  In order to do so, they are given the
additional resources of synchronized clocks and perfect local control.
They may attach or discard local ancillary systems and they may apply
(arbitrarily fast) local operations, which can be assumed to be
unitary without loss of generality \cite{BCL02}.
As a side remark, classical communication between Alice and Bob would
not increase the optimal simulation rate \cite{VC02}.
Note that the goal of the simulation is not to produce the Hamiltonian
evolution $e^{-i H' t}$ for a particular time $t$, but rather to
stroboscopically track the evolution $e^{-i H' t}$ for arbitrarily
closely spaced values of time.
A detailed formulation of the problem can be found in \cite{BCL02}. 

We next present a list of rules for nonlocal Hamiltonian simulation.
By composition, these rules give rise to all possible simulations
achievable with local operations and ancillary systems.
We present five basic rules, as well as three additional rules that
can be obtained by combining the basic ones.
We use the shorthand notation $H \longrightarrow H'$ to represent the
possibility of simulating $H'$ by $H$ at the rate $\g_{H'|H}$, and the
notation $H \longleftrightarrow H'$ to indicate that, in addition, the
simulation can be reversed without overall efficiency losses, as in
(\ref{eq:reversible}).  We say that two Hamiltonians are {\em locally
equivalent} if they can simulate each other reversibly {\em at unit
rate}.

The first two basic rules merely make precise the notion of
Hamiltonian evolution.  They do not involve any operational procedure,
nor assume any ability to control the system.
The first rule makes precise the notion of rescaling the evolution
time: a Hamiltonian $H$ can reversibly simulate another Hamiltonian
$H'=cH$ that only differs by a {\em positive} multiplicative constant
$c$.
\begin{arule}[Rescaling]
\label{rule:scale}
  For any $c>0$,
  \be
    H \longleftrightarrow cH\,, \qquad
    \g_{cH | H} = {1 \over c}
  \,.
  \ee
\end{arule}
\noindent
Note that it is important that $c>0$.  In general, Hamiltonians $H$
and $-H$ cannot simulate each other reversibly (see
\cite{BCL02,WRJB02} for examples).

The second rule makes precise what it means for a Hamiltonian to act
on a subsystem.  In the bipartite setting, the complete system can be
described by subsystems $A,B$ on which $H$ acts and ancillary
subsystems $A',B'$ on which it acts trivially.
\begin{arule}[Ancillas]
\label{rule:attach}
 For any dimension of the ancillary Hilbert space 
  $\H_{A'} \otimes \H_{B'}$,
  \be
    H \longleftrightarrow  H \otimes I_{A'B'}\,, \qquad
    \g_{H \otimes I_{A'B'} | H} = 1
  \,.
  \ee
\end{arule}

The next two basic rules arise from the possibility of switching on
local Hamiltonians.
\begin{arule}[Local Hamiltonians]
\label{rule:local}
  Any local Hamiltonian of the form $H_0 = H_{A} \otimes I_{B} +
  I_{A} \otimes H_{B}$ can be produced at no cost.
\end{arule}
Recall that for any unitary transformation $U$, we have $U e^{-iHt}
U^\dagger = e^{-i UHU^\dagger t}$.  Therefore, by means of local
unitaries, any Hamiltonian $H$ is locally equivalent to any other 
that is obtained from it by local unitary conjugation.
\begin{arule}[Local unitaries]
\label{rule:unitary}
  For any local unitary operation $U=U_{A} \otimes U_{B}$,
  \be
    H \longleftrightarrow UHU^{\dagger}\,, \qquad 
    \g_{U H U^\dag | H} = 1
  \,.
  \ee
\end{arule}

Rules \ref{rule:scale}--\ref{rule:unitary} allow Alice and Bob to
produce Hamiltonians that differ from the original interaction $H$.
The Lie-Trotter product formula, 
\be
  e^{-i(H_1+H_2)t} = 
  \lim_{n \to \infty} (e^{-i H_1 t/n} e^{-i H_2 t/n})^n
\,,
\ee
tell us how two of these Hamiltonians can be combined into a new one
by alternately simulating each of them individually.  With the help of
Rule \ref{rule:scale} we obtain the last basic rule.
\begin{arule}[Convex combination]
\label{rule:sum}
  For any $H_1, H_2$ and $0\leq p \leq 1$, the simulation
  \be
    \left. \begin{array}{rr}
    p \, H_1 \\
    (1-p) H_2 
    \end{array} \right\}
    \longrightarrow H' = pH_1 + (1-p)H_2
  \label{eq:psum}
  \ee
  is possible with rate $\g_{H'| pH_1; (1-p)H_2} \ge 1$.
\end{arule}
\noindent
Here we have considered the use of Hamiltonian $H_1$ for a fraction of
time $p$ and Hamiltonian $H_2$ for a fraction of time $1-p$, and the
rate of simulating $H'$ is computed by adding these two times
together.  
Let us stress that (\ref{eq:psum}) assumes only the local ability to
switch on and off the constituent Hamiltonians, and that only one
Hamiltonian is acting at a time.
Notice that this is the only basic simulation rule where
irreversibility may occur.  Although Alice and Bob can use $H'$ to
simulate back $H_1$ and $H_2$, in general they will incur an overall
loss in efficiency by doing so.

These basic rules can be combined in various ways.  We state three
particularly useful combinations as additional rules.  
First, from Rules \ref{rule:local} and \ref{rule:sum}, a local part
$H_0$ can be added to the given nonlocal Hamiltonian reversibly.
\begin{arule}[Adding a local Hamiltonian]
\label{rule:localsum}
\be
  H \longleftrightarrow H + H_0 \,, \qquad 
  \g_{H+H_0 | H} = 1
\,.
\ee
\end{arule}
Second, local unitary conjugation and convex combination can be composed
into what we shall call a local unitary mixing of $H$.
\begin{arule}[Local unitary mixing]
\label{rule:unitarymixing}
For any set of local unitary transformations $U_i = U_{A,i}\otimes
U_{B,i}$ and any probability distribution $p_i$ ($p_i \ge 0$ and
$\sum_i p_i = 1$),
\be
  H \longrightarrow H' = \sum_i p_i U_i H U_i^\dag \,, \qquad
  \g_{H' | H} \ge 1
\,.
\ee
\end{arule}

Note that Rules \ref{rule:local}--\ref{rule:unitarymixing} are stated
without assuming local control over the ancillas.  
In the two-qubit case, Rules \ref{rule:scale}, \ref{rule:localsum},
and \ref{rule:unitarymixing} describe all relevant simulations because
local control over ancillas is known to be unnecessary for optimal
simulations \cite{BCL02}.

More generally, we allow local control over ancillas in our simulation
model.  By Rule \ref{rule:attach}, Rules
\ref{rule:local}--\ref{rule:unitarymixing} can be extended to include
ancillas as well.
Control over ancillas gives extra freedom in the simulation, and is
known to improve the achievable simulation rates in some cases
\cite{VC02}.

Our last rule is concerned with any simulation in which the original
Hamiltonian $H$ and the simulated Hamiltonian $H'$ act on systems with
different dimensions.  Let $\H_A \otimes \H_B$ and $\H_{A'}\otimes
\H_{B'}$ denote the Hilbert spaces on which $H$ and $H'$ act, with
dimensions $d_A,d_B$ and $d_{A'}, d_{B'}$, where $d_A \geq d_{A'}$ and
$d_B \geq d_{B'}$.  For simplicity, we assume that $H=H_A\otimes H_B$
is a product Hamiltonian. If it were not, then we could expand $H$ as
a linear combination of product Hamiltonians, $H=\sum_i H_{A,i}\otimes
H_{B,i}$, and the following would hold for each of the terms in the
expansion.  Let vectors $|j\>_A$ ($1\leq j\leq d_A$) denote an
orthonormal basis in $\H_A$. We can express $H_A$ as
\be
  H_{A} = \left[ \begin{array}{cc} 
    J_\parallel & C^\dag \\
    C           & J_\perp
  \end{array} \right]
\,,
\ee
where $J_{\parallel}$ is the restriction of $H_A$ onto the subspace
$\H_{A\parallel} \subseteq \H_A$ spanned by vectors $|j\>_A$ ($1\leq j\leq
d_{A'}$), and $J_{\perp}$ the projection onto its orthogonal
complement.  Consider also an analogous decomposition for $H_B$.  Then
we have the following:

\begin{arule}[Reduction to a local subspace]
\label{rule:reduction}
The simulation
\be
  H = \left[ \begin{array}{cc} 
    J_{\parallel} & C^{\dagger}\\
    C & J_{\perp}
  \end{array} \right]
  \otimes \left[ \begin{array}{cc} 
    K_{\parallel} & D^{\dagger}\\
    D & K_{\perp}
  \end{array} \right] \longrightarrow
  H' = J_{\parallel}\otimes K_{\parallel}
\ee
is possible with rate $\g_{H' | H} \ge 1$.
\end{arule}

The last rule can be obtained by using the following lemma twice.  The
lemma shows how to simulate $J_{\parallel}\otimes H_B$ using
$H_A\otimes H_B$.

\begin{lemma}
The simulation
\be
  H = \left[ \begin{array}{cc} 
    J_{\parallel} & C^{\dagger}\\
    C & J_{\perp}
  \end{array} \right] \otimes H_B
  \longrightarrow H' = J_{\parallel}\otimes H_B
\ee
is possible with rate $\g_{H' | H} \ge 1$.
\end{lemma}

\begin{proof}
We divide the simulation into two steps. 
($i$) First, by unitary mixing (Rule \ref{rule:unitarymixing}) with
\bea
  p_1 = \frac{1}{2},~~~ U_1 &=& I_A\otimes I_B; \nonumber \\ 
  p_2 = \frac{1}{2},~~~ U_2 &=& \left[ \begin{array}{cc} 
                                I_{\parallel} & 0 \\
                                0 & -I_{\perp}
  \end{array} \right] \otimes I_B
\,,
\eea
where $I_\parallel$ and $I_\perp$ denote restrictions of the identity
operator, we achieve the simulation 
\be
  H = \left[ \begin{array}{cc} 
    J_{\parallel} & C^{\dagger}\\
    C & J_{\perp}
  \end{array} \right] \otimes H_B 
  \longrightarrow 
  H'' = \left[ \begin{array}{cc} 
    J_{\parallel} & 0\\
    0 & J_{\perp}
  \end{array} \right] \otimes H_B
\ee
with unit rate, so $\g_{H''|H} \ge 1$.  
($ii$) Second, we use $H''$ to simulate $H'= J_{\parallel} \otimes
H_B$ as follows.  Suppose the goal is to evolve $|\psi\>_{A'B}$
according to $e^{-iH't}$.
We assume system $A$ is in state $\ket{1}_{A}$ by local control.
Therefore the joint state of systems $AA'B$ is initially
$\ket{1}_{A}\ket{\psi}_{A'B}$.  Let $V_{AA'}$ denote a unitary
transformation such that
\be
  V_{AA'} \ket{1}_{A}\ket{j}_{A'} =  \ket{j}_{A}\ket{1}_{A'} \,, \quad
  1\leq j\leq d_{A'} 
\,.
\ee 
Then the following three steps can be used to complete the desired
simulation:
\begin{enumerate}
\item We apply unitary $V_{AA'}$, placing $\ket{\psi}_{A'B}$ in the
      subspace $\H_{A\parallel}\otimes \H_B \subset \H_A \otimes
      \H_B$.
\item We make $AB$ evolve according to $H''$. Notice that at all times
      $e^{-iH''t}\ket{\psi}_{AB}$ is supported in $\H_{A\parallel}
      \otimes \H_B$, and that $H''$ acts on this subspace as
      $J_{\parallel}\otimes H_B$.
\item We reverse $V_{AA'}$, so that the net evolution on $A'B$ has been
      $e^{-iH't}\ket{\psi}_{A'B}$.
\end{enumerate}
This completes the proof.
\end{proof}

\section{Reversible simulation}
\label{sec:equiv}

In this section, we present the main result of the paper, the
reversible simulation of tensor product Hamiltonians $H = H_A \otimes
H_B$. We will consider product Hamiltonians in a certain standard
form. Using Rule \ref{rule:unitary}, we may diagonalize $H_A$ and
$H_B$, so we need only consider their eigenvalues.  It will also be
convenient to modify $H_A$ so that the largest and smallest
eigenvalues, $\lambda^{\rm max}_A$ and $\lambda^{\rm min}_A$, are equal
in magnitude, and similarly for $H_B$.  This can be done by adding a
term proportional to the identity to each of $H_A$ and $H_B$, i.e.,
\bea
     \!\!\!\!\!\!\!\!
  (H_A + c I) \otimes (H_B + d I) = 
  && \!\!\!\!\!\!        H_A \otimes H_B + c   \, I \otimes H_B \nn
  && \!\!\!\!\!\! + \, d \, H_A \otimes I   + c d \, I \otimes I
\,.
\eea
The resulting Hamiltonian is locally equivalent to $H$ since they
differ only by local terms (Rule \ref{rule:localsum}).  Furthermore,
since
\be
 (c H_A) \otimes (H_B/c) = H_A \otimes H_B
\,,
\ee
we may assume $\lambda^{\rm max}_A - \lambda^{\rm min}_A =
\lambda^{\rm max}_B - \lambda^{\rm min}_B = \Delta$ without loss of
generality.

Having put all product Hamiltonians into a standard form, we are ready
to show that they can reversibly simulate each other.  
By the transitivity of reversible simulation, it suffices to show that
all product Hamiltonians can reversibly simulate the Ising interaction
$H_{\rm Ising} = \s{z} \otimes \s{z}$.
\begin{lemma}
\label{lem:productIsing}
Any tensor product Hamiltonian $H=H_A\otimes H_B$ can reversibly
simulate the Ising interaction
\be
  H \longleftrightarrow H_{\rm Ising} \,, \quad 
  \g_{H_{\rm Ising}|H} = \frac{1}{4} \Delta^2
\,.
\ee
\end{lemma}

\begin{proof}
For any nonlocal $H_1$ and $H_2$, we have
\be
  \g_{H_1|H_2} \g_{H_2|H_1} \leq 1 \,.  
\ee
Otherwise we could use $H_1$ to simulate itself with simulation rate
greater than 1, which is a contradiction.\footnote{If $\g_{H_1|H_2}
  \g_{H_2|H_1} > 1$, then we could concatenate several simulations
  $H_1\longrightarrow H_2\longrightarrow H_1\longrightarrow \cdots
  \longrightarrow H_2\longrightarrow H_1 $ to obtain that the optimal
  simulation rate $\g_{H_1|H_1}$ is infinite. Recalling that any
  bipartite nonlocal Hamiltonian can simulate any other one at finite
  rate, we would conclude that $\g_{H|H}$ is also infinite for any
  bipartite Hamiltonian. This would contradict, for instance, the
  results of \cite{BCL02} showing that $\g_{H|H}=1$ for all nonlocal
  two-qubit Hamiltonians.}
It thus suffices to show that $\g_{H_{\rm Ising}|H} \ge \Delta^2/4$
and $\g_{H|H_{\rm Ising}} \ge 4/\Delta^2$.

Let $H$ act on a Hilbert space $\H_A \otimes \H_B$ with dimensions
$d_A$ and $d_B$. Since $H=H_A \otimes H_B$ is in the standard form, we
may write
\bea
  H = & \! {1 \over 4} \Delta^2 \!\! 
      & \diag(1,a_2,a_3,\ldots,a_{d_{\!A}\!-\!1},-1) \nn
      & \otimes \!\!\!\!\!\!\!\!     
      & \diag(1,b_2,b_3,\ldots,b_{d_{\!B}\!-\!1},-1)
\eea
where
\bea
  1 = a_1 \geq a_2 \geq &\ldots& \geq a_{d_A} = -1 \\
  1 = b_1 \geq b_2 \geq &\ldots& \geq b_{d_B} = -1
\,,
\eea
and the corresponding eigenvectors are $|j\>_A$ ($1 \le j \le d_A$) and
$|j\>_B$ ($1 \le j \le d_B$).  

We can simulate the Ising interaction using $H$ by restricting to the
subspace spanned by the extremal eigenvectors of $H_A$ and $H_B$
($|1\>_A |1\>_B$, $|d_A\>_A |1\>_B$, $|1\>_A |d_B\>_B$, $|d_A\>_A
|d_B\>_B$) according to Rule \ref{rule:reduction}.
In this subspace, $H$ acts as $(\Delta^2/4) H_{\rm Ising}$.  Therefore
we have
\be
  \g_{H_{\rm Ising}|H} \ge \Delta^2/4
\,.
\label{eq:hising}
\ee

In order to show how to use the Ising interaction $H_{\rm Ising}$ to
simulate $H$, we consider a concatenation of two simulations, 
\be
  H_{\rm Ising} \longrightarrow H'' \longrightarrow H
\,.
\ee
Here $H''= {1 \over 4} \Delta^2 H''_{AA'} \otimes H''_{BB'}$ acts on 
local Hilbert spaces of dimensions $2d_A$ and $2d_B$, and reads 
\bea
  H'' = & \! {1 \over 4} \Delta^2 \!\! 
        & \diag(1,-1,a_2,-a_2,\cdots,-1,1) \nn
        & \otimes \!\!\!\!\!\!\!\!
        & \diag(1,-1, b_2,-b_2,\ldots,-1,1)
\,.
\eea
Clearly, we can use Rule \ref{rule:reduction} to simulate $H$ by $H''$
with unit simulation rate. Therefore we need only focus on the
simulation of $H''$ by $H_{\rm Ising}$. In turn, this can be decomposed
into two similar simulations, 
\be
  H_{\rm Ising} \longrightarrow H''_{AA'} \otimes \s{z}
                \longrightarrow H''_{AA'} \otimes H''_{BB'}
\,,
\label{eq:double}
\ee 
each one with unit rate. In order to simulate $H''_{AA'} \otimes
\sigma_z$ using $\sigma_z \otimes \sigma_z$, we append a
$d_A$-dimensional ancilla $A$ to qubit $A'$ (with $H_{\rm Ising}$
acting on $A'B'$) to obtain the Hamiltonian
\bea
  H_{\rm Ising}' &=& (I\otimes \s{z,A'}) \otimes \s{z} \\
                 &=& \diag(1,-1,1,-1,\ldots,1,-1) \otimes \s{z}
\,.
\eea
We define
\be
  p_j = (a_j+1)/2
\label{eq:a2p} 
\ee
so that
\be
  1 = p_1 \geq p_2 \geq \ldots \geq p_{d_A} = 0
\,.
\ee
Furthermore, we define $d_A$ local unitary operations
$\{U_j\}_{j=1}^{d_A}$, where $U_j$ exchanges the $(2j\!-\!1)$-th and
$(2j)$-th basis vectors of $AA'$.  To evolve under $H''_{AA'} \otimes
\s{z}$ for a small time $\delta$, we apply each $U_j$ at time $t=p_j
\delta$ and $U_j^\dag$ at time $t=\delta$.  Equivalently, we can use
Rule \ref{rule:unitarymixing} with an appropriate probability
distribution and set of unitaries.  Thus we can use $H_{\rm Ising}$ to
simulate $H''_{AA'} \otimes \s{z}$ with unit efficiency.  The second
simulation in (\ref{eq:double}) is achieved similarly.  The overall
rate for $H_{\rm Ising}$ to simulate $H''$ or $H$ is thus $4/\Delta^2$
by Rule \ref{rule:scale}.
\end{proof}

We have shown that any product Hamiltonian $H$ can reversibly simulate
the Ising interaction $H_{\rm Ising}$ with rate $\g_{H_{\rm Ising}|H}=
K_{\otimes}(H)$, where
\be
  K_{\otimes}(H)
  = {1 \over 4}\Delta_A \Delta_B
\,.
\ee 
Our main result then follows: 

\begin{theorem}
\label{thm:product}
Any product Hamiltonian $H$ can reversibly simulate any other product
Hamiltonian $H'$, with simulation rate given by
\be
  \g_{H'|H} = \frac{K_{\otimes}(H)}{K_{\otimes}(H')}.
\ee
\end{theorem}

As discussed previously, in general a bipartite Hamiltonian $H$ cannot
reversibly simulate $-H$.  Similarly, in general $H$ cannot reversibly
simulate its complex conjugate $H^*$, nor the Hamiltonian $H^\diamond$
resulting from swapping systems $A$ and $B$.  However, for product
Hamiltonians, all these Hamiltonians are locally
equivalent.

\begin{corollary}
\label{coro:inverse}
For any product Hamiltonian $H$,
\be
  K_\otimes(H) = K_\otimes(-H)
               = K_\otimes(H^*)
               = K_\otimes(H^\diamond)
\,.
\ee
\end{corollary}

Theorem \ref{thm:product} can be extended to the case of a sum of
bipartite product Hamiltonians acting on separate systems.  If $H_1$
and $H_2$ are two Hamiltonians acting, respectively, on bipartite
systems $A_1B_1$ and $A_2B_2$, we let $H_1 \boxplus H_2$ denote their
sum.\footnote{We use the symbol $\boxplus$ rather than $+$ to
  emphasize that the Hamiltonians being summed act on different pairs
  of systems.  In other words, $H_{AB} \boxplus H_{A'B'} = H_{AB}
  \otimes I_{A'B'} + I_{AB} \otimes H_{A'B'}$.}
In fact, $H_1 \boxplus H_2$ can reversibly simulate a product
Hamiltonian $H$ acting on a single bipartite system $AB$.

\begin{lemma} 
\label{lm:three}
If $H_1$, $H_2$, and $H'$ are product Hamiltonians, the simulation
\be
  H_1 \boxplus H_2 \longleftrightarrow H'
\ee
can be achieved reversibly, with simulation rate
\be
  \g_{H'|H_1 \boxplus H_2} = \g_{H'|H_1}+\g_{H'|H_2}
\,.
\ee
\end{lemma}

\begin{proof}
Because of Theorem \ref{thm:product}, we only need to show that the
Hamiltonian $H'= c \, H_{\rm Ising} \boxplus d \, H_{\rm Ising}$, $c,d
\in \R$, can reversibly simulate $H=(|c|+|d|)H_{\rm Ising}$ at unit
rate.  In addition, Corollary \ref{coro:inverse} implies that we need
only consider the case $c,d > 0$.

By Rule \ref{rule:attach}, $H = (c+d) H_{\rm Ising}$ is locally equivalent
to
\be
  J_1 = (c + d) \, \s{zA} \otimes \s{zB} \otimes I_{A'B'}
\,.
\ee
In turn, using local unitaries to swap $A$ with $A'$ and $B$ with $B'$
(Rule \ref{rule:unitary}), $J_1$ is locally equivalent to
\be
  J_2 = (c + d) \, I_{AB} \otimes \s{zA'} \otimes \s{zB'}
\,.
\ee
Then we can simulate $H' = [c / (c + d)] J_1 + [d/(c + d)] J_2$ by
convex combination (Rule \ref{rule:sum}) of $J_1$ and $J_2$, which
shows that $\g_{H'|H} \ge 1$.

For the reverse simulation, note that $H'$ is locally equivalent to each
of the following four Hamiltonians:
\be
\begin{array}{@{c(}        c@{\otimes}c@{\otimes}c@{\otimes}c@{)}
              @{\; + \; d(}c@{\otimes}c@{\otimes}c@{\otimes}c@{)}l}
 \s{zA} & I_{A'}  & \s{zB} & I_{B'}  & I_{A}  & \s{zA'} & I_{B}  &\s{zB'} &\,,\\
 I_{A}  & \s{zA'} & \s{zB} & I_{B'}  & \s{zA} & I_{A'}  & I_{B}  &\s{zB'} &\,,\\
 \s{zA} & I_{A'}  & I_{B}  & \s{zB'} & I_{A}  & \s{zA'} & \s{zB} & I_{B'} &\,,\\
 I_{A}  & \s{zA'} & I_{B}  & \s{zB'} & \s{zA} & I_{A'}  & \s{zB} & I_{B'} &\,.
\end{array}
\ee
Each of these Hamiltonians can be obtained from $H'$ according to Rule
\ref{rule:unitary} by swapping $A$ with $A'$ and $B$ with $B'$ as
necessary. An equally weighted convex combination of these four
Hamiltonians gives, after rearranging terms,
\be
  {c \! + \! d \over 4} \; 
  (\s{zA} \!\otimes\! I_{A'} + I_{A} \!\otimes\! \s{A'}) 
  \otimes (\s{zB} \!\otimes\! I_{B'} + I_{B} \!\otimes \!\s{B'}) 
\,, 
\ee
a tensor product Hamiltonian with $\Delta^2 = 4 \, (c + d)$.  Therefore
$\g_{H|H'} \ge 1$, which completes the proof.
\end{proof}

It follows from Lemma \ref{lm:three} that $K_{\otimes}$ is {\em additive}
under the sum of product Hamiltonians acting on different pairs of
systems,
\be
  K_\otimes(H_1 \boxplus H_2) = K_{\otimes} (H_1) + K_{\otimes} (H_2) 
\,.
\ee
More generally, for $H = \bigboxplus_i H_i$ and $H' = \bigboxplus_i H'_i$,
where all $H_i$ and $H_i'$ are bipartite product Hamiltonians, we can
perform the simulation
\be
  H = \bigboxplus_i H_i \longleftrightarrow H' = \bigboxplus_i H'_i
\ee
reversibly, with simulation rate given by
\be
  \g_{H'|H} = \frac{\sum_i K_\otimes (H_i)}{\sum_i K_\otimes (H'_i)}
\,.
\ee

Finally, in the Appendix, we present a case of reversible Hamiltonian
simulation that is possible when in addition to local operations and
ancillas, catalytic pre-shared entanglement is available to Alice and
Bob.  The simulation can be made reversible only in the presence of
entanglement, but the entanglement is not used up during the
simulation \cite{VC02a}.

\section{Entanglement capacity}
\label{sec:ent}

In this section we use the results on reversible simulation from
Sec.~\ref{sec:equiv} to determine the optimal way a product
Hamiltonian can be used to produce entanglement.

The problem of optimal entanglement generation by an interaction
Hamiltonian has been approached in different ways.  In \cite{DVCLP01}
the {\em single-shot} scenario was considered.  This corresponds to
the setting in which an interaction $H$ is used, with the help of fast
local unitary transformations (but without control over local
ancillas), to maximize the rate of increase of entanglement between a
single copy of the two interacting systems.  This situation is of
interest for many present day experiments aiming to produce entangled
states in quantum optics, nuclear magnetic resonance, or condensed
matter systems \cite{FortPhys}.  The optimal rate at which
entanglement can be generated in the single-shot scenario is known as
the {\em entanglement capability} of $H$, $\Gamma_H$, and its value
has been determined for any two-qubit Hamiltonian \cite{DVCLP01}.  For
the Ising interaction it reads
\be
  \Gamma_{H_{\rm Ising}} = \alpha \approx 1.9123
\,,
\ee
where
\be
  \alpha := 2 \max_{x \in [0,1]} \sqrt{x (1-x)}
            \log\left({x \over 1-x}\right)
\ee
with the maximum obtained at $x_0 \approx 0.9168$.

In contrast, \cite{BHLS02} considers entanglement generation in the
{\em asymptotic} scenario, where many copies of the interacting
systems (and local ancillas) can be used collectively to produce
entanglement, possibly at a higher rate than in the single-shot case.
The {\em asymptotic entanglement capacity} (or {\em entanglement
capacity} for short) of interaction $H$, denoted $E_{H}$, is of
interest in the context of understanding the ultimate limitations of
quantum mechanical systems to process quantum information.  Reference
\cite{BHLS02} showed that entanglement capacities equal entanglement
capabilities when ancillas are allowed in the single-shot setting. 

Finally, it was shown in \cite{CLVV03} that for a number of two-qubit
Hamiltonians, all single-shot and asymptotic scenarios lead to the
same optimal entanglement generation rates.  In particular,
\be
  E_{H_{\rm Ising}} = \Gamma_{H_{\rm Ising}} = \alpha
\,.
\ee

Combining (\ref{eq:rev}) and Theorem \ref{thm:product}, we obtain an
expression for the entanglement capacity of any product Hamiltonian
$H$, since we have $E_{H} = K_{\otimes}(H) \, E_{H_{\rm Ising}}$.
\begin{theorem}
\label{thm:entcap}
For any product Hamiltonian $H$,
\be
  E_{H} = \frac{\alpha}{4}\Delta_A\Delta_B
\,.
\label{eq:EH}
\ee
\end{theorem}
\noindent 
Similarly, we can compute $E_{H_1 \boxplus H_2}$ for the sum of two
product Hamiltonians $H_1$ and $H_2$ acting on different pairs of
systems:
\be
  E_{H_1 \boxplus H_2} = E_{H_1} + E_{H_2}
\,.
\label{eq:EH12}
\ee

In fact, (\ref{eq:EH}) also corresponds to the single-shot capability
$\Gamma_H$, since it can be obtained without using ancillas.
\begin{corollary}
\label{cor:entcap}
For any product Hamiltonian $H$,
\be
  \Gamma_H = {\alpha \over 4} \Delta_A\Delta_B
\,.
\ee
\end{corollary}

\begin{proof}
The explicit optimal input state is
\be
  |\psi\> = \sqrt{x_0}   |+\>_A \otimes |+\>_B
        + i \sqrt{1-x_0} |-\>_A \otimes |-\>_B
\ee
where
\be
  |\pm\>_A = {1 \over \sqrt 2} (|1\>_A \pm |d_A\>_A)
\ee
and similarly for system $B$.  Here $|1\>_A$ and $|d_A\>_A$ represent
the eigenstates of $H_A$ corresponding to the largest and smallest
eigenvalues, respectively.  That this state achieves $\Gamma_H$ can be
seen by substitution into Eqs.~(17) and (18) of \cite{CLVV03}.
\end{proof}
\noindent Likewise, (\ref{eq:EH12}) can also be achieved without
ancillas, because the protocol used in Lemma \ref{lm:three} does 
not involve ancillas: 
\be
  \Gamma_{H_1 \boxplus H_2} = \Gamma_{H_1} + \Gamma_{H_2}
\,.
\ee

As a side remark, \cite{WS02} has reported the restricted case of
(\ref{eq:EH}) in which $H_A$ and $H_B$ have eigenvalues $\pm 1$, but
this property has no special significance; only the tensor product
structure is important.

Equation (\ref{eq:EH}) could also be proved directly using essentially
the same arguments that appear in \cite{CLVV03}.  Using the
observation that the matrix with elements
\be
  4|(H_A)_{jk} (H_B)_{jk}|/\Delta^2
\ee
(where the matrix elements of $H_A$ and $H_B$ are taken in any
orthonormal basis) is doubly substochastic, an upper bound on $E_H$
can be obtained just as in Eq.~(26) of \cite{CLVV03}.  However, the
approach using reversible simulation is more powerful, since it
applies to any capacity, even those for which the capacity of $H_{\rm
Ising}$ is yet to be found, such as the capacity for communicating
classical information.

Finally, we note that Theorem \ref{thm:entcap} can be extended to
Hamiltonians that can be reversibly simulated using catalytic
entanglement, such as those mentioned in Sec.~\ref{sec:equiv} and
further described in the Appendix.  
This class of Hamiltonians includes, as a special case, the full set
of two-qubit Hamiltonians of the form $\mu_x \, \s{x} \otimes \s{x} +
\mu_y \, \s{y} \otimes \s{y}$ considered in \cite{CLVV03}.
In the context of asymptotic entanglement capacity, catalytic
resources need not be considered as additional requirements since the
cost of first obtaining any catalytic resource can be made negligible
\cite{BHLS02}.
However, it turns out that for the Hamiltonians discussed in the
Appendix, catalytic entanglement is actually not necessary to achieve
the entanglement capacity.

\section{Strength measures for bipartite Hamiltonians}
\label{sec:strength}

In this section, we show that the notion of nonlocal Hamiltonian
simulation can be used to define measures of the strength of a
nonlocal Hamiltonian.  We proceed along the lines of \cite{NDD02},
which introduces a formalism of {\em strength measures} for nonlocal
quantum operations, with the goal of quantifying their nonlocality.
Three necessary axioms and a number of other desirable properties for
such measures were proposed.  Here we show that in the case of
nonlocal Hamiltonians, a single axiom implies many of these
properties.

We denote a strength measure for Hamiltonian $H$ as $K(H)$. The only
requirement we impose on $K(H)$ is that it does not increase under
Hamiltonian simulation by local manipulations.
That is, if $H$ can be used to simulate $H'$ at a rate $\g_{H'|H}$,
then the strength measure for $\g_{H'|H}\, H'$ should be no greater
than that for $H$. 
This is motivated by the idea that any measure of the {\em
nonlocality} of an interaction should not increase under {\em local}
manipulations.

\begin{axiom}[Monotonicity] Any strength measure $K(H)$ must satisfy
\be
  K(H) \ge \g_{H'|H} K(H')
\label{eq:axiom}
\ee 
for any two Hamiltonians $H,H'$.\footnote{
  One might also be interested in defining a strength measure in a
  setting where more than just a single Hamiltonian is available.
  Suppose we can use two switchable Hamiltonians $H_1$ and $H_2$ to
  achieve some task, and that we can only have one Hamiltonian
  switched on at a time.  Let $p$ and $1-p$ ($0\leq p \leq 1$)
  characterize the relative frequency with which we use each of the
  two Hamiltonians.  Then we can define a function $K(pH_1;(1-p)H_2)$
  as a strength measure. More generally, $K$ may have any number of
  arguments.  Suppose we can use $n$ Hamiltonians $\{p_1H_1; p_2H_2;
  \cdots; p_n H_n\}$, where $p_i$ ($p_i\geq 0$, $\sum_ip_i=1$)
  indicates the relative frequency of $H_i$, to simulate Hamiltonians
  $\{q_1H'_1; q_2H'_2; \cdots; q_m H'_m\}$.  Let
  $\g_{\{q_iH'_i\}|\{p_iH_i\}}$ denote the rate at which this can be
  achieved. Then we would require that a strength measure $K$ fulfills
  $K(\{p_iH_i\}) \geq \g_{\{q_iH'_i\}|\{p_iH_i\}} K(\{q_iH'_i\})$.
  Several additional properties can be derived for $K$.  In
  particular, Rule \ref{rule:sum} implies that $K(\{p_iH_i\}) \geq
  K(\sum_i p_iH_i)$.} 
\end{axiom}

In the following, we exclude the trivial strength measure $K(H)=0$ for
all $H$.  Likewise, we exclude any unphysical strength measure that is
infinite for a bounded $H$.  

We have already encountered two examples of non-trivial functions that
do not increase under local manipulations: $(i)$ any capacity $C_H$
satisfies the Axiom because of (\ref{eq:simbound}), and ($ii$) for any
fixed target Hamiltonian $H_{\rm target}$, the simulation rate
$\g_{H_{\rm target}|H}$ satisfies the Axiom as a function of $H$. In
particular, the function
\be
  K_{\otimes} (H) := \g_{H_{\rm Ising}|H}
\ee
used in previous sections is a strength measure.

A function $K(H)$ satisfying the Axiom automatically has a number of
properties that we describe below for the bipartite case (although
much of the discussion can be generalized to more than two systems).
Several of these properties appeared in the original formulation in 
\cite{NDD02}.

\begin{prop}[Positivity] 
\label{prop:pos}
\be
  K(H) \ge 0
\ee
with equality if and only if $H$ is local.
\end{prop}

\begin{prop}[Homogeneity] 
\label{prop:hom}
For any $c \ge 0$,
\be
  K(c H) = c K(H)
\,.
\ee
\end{prop}

\begin{prop}[Stability under addition of ancillas] 
\label{prop:sta}
For any ancillary system $A'B'$,
\be
  K(H) = K(H\otimes I_{A'B'}).
\ee
\end{prop}

\begin{prop}[Local unitary invariance] 
For any local unitary operation $U=U_A\otimes U_B$,
\label{prop:loc}
\be
  K(H) = K(UHU^{\dagger})
\,.
\ee
\end{prop}

\begin{prop}[Invariance under local addition]
For any local Hamiltonian $H_0$,
\label{prop:inv}
\be
  K(H) = K(H+H_0)
\,.
\ee
\end{prop}

\begin{prop}[Local unitary mixing]
\label{prop:unimix}
For any set of local unitary operators $U_i=U_{A,i}\otimes U_{B,i}$
and probability distribution $p_i$,
\be
  K(H) \geq K\left(\sum_i p_i U_i H U_{i}^{\dagger}\right)
\,.
\ee
\end{prop}

\begin{prop}[Reduction to a local subspace]
\label{prop:red}
Let $J_{\parallel}$ denote the $m\times m$ upper left submatrix of $J$,
and $K_{\parallel}$ the $m'\times m'$ upper left submatrix of $K$. Let
$H=\sum_i J_i\otimes K_i$ be a product expansion of a bipartite
Hamiltonian $H$, and let $H'=\sum_i J_{i\parallel} \otimes K_{i
\parallel}$ be its restriction to the upper left subspace.  Then
\be
  K(H) \ge K(H')
\,.
\ee
\end{prop}

\begin{prop}[Continuity]
For any nonlocal Hamiltonian $H$ and any bounded Hermitian operator $J$,
\label{prop:cont}
\be
  \lim_{\epsilon \to 0} K(H+\epsilon J)-K(H) = 0
\,.
\ee
\end{prop}

We now derive Properties \ref{prop:pos}--\ref{prop:cont} from the
Axiom and our simulation rules in Section \ref{sec:sim}.  First we
prove Property \ref{prop:pos}:
\begin{proof}
Let $H_0$ be any local Hamiltonian and $H$ be any bounded nonlocal
Hamiltonian.  $H_0$ cannot simulate $H$, so $\g_{H|H_0}=0$.  The Axiom
then implies $K(H_0) \geq 0$.  Through local control, $H_0$ can be
simulated without using $H$, so $\g_{H_0|H} > 0$.  The Axiom then
implies $K(H) \geq 0$.  However, $\g_{H_0|H}$ is unbounded.  If $K(H)$
is bounded, $K(H_0)$ must be $0$.  Finally, if $K(H) = 0$, the fact
that it can simulate all other (bounded) Hamiltonians $H'$ with
nonzero rate implies $K(H') = 0$, which is excluded.  Thus $K(H)>0$.
\end{proof}
\noindent
Properties \ref{prop:hom}--\ref{prop:red} follow from the simulations
described, respectively, in Rules \ref{rule:scale}, \ref{rule:attach},
\ref{rule:unitary}, \ref{rule:localsum}--\ref{rule:reduction}.
Finally, the proof of Property \ref{prop:cont} is based on the fact
that any two nonlocal bipartite Hamiltonians can simulate each other
with nonzero rate:
\begin{proof}
Because nonlocal strength measures are invariant under addition of
local terms (Property \ref{prop:inv}), we can focus on a Hamiltonian
$H$ without local terms, that is, $\tr_{A} H=\tr_B H = 0$, and
similarly for $J$.  Consider first the case where $J$ is proportional
to $H$.  Then Property \ref{prop:hom} ensures continuity.  Suppose now
that $J$ is not proportional to $H$, so that $H+\epsilon J \ne 0$
$\forall \epsilon$ and is therefore always nonlocal.  We then consider
the following two simulations:
\bea
  && H \longrightarrow H + \epsilon J \label{eq:forth}\\
  && H + \epsilon J \longrightarrow H \label{eq:back}
\,.
\eea
A possible procedure for (\ref{eq:forth}) is to use $H$ to simulate
itself at rate $\g_{H|H}$ and $\epsilon J$ at rate $\g_{\epsilon
J|H}$, and to add the two Hamiltonians together by Rule
\ref{rule:sum}.  This gives a lower bound on the optimal simulation
rate
\be
  \g_{H + \epsilon J|H} \geq
    \left({1 \over \g_{H|H}} + {1 \over \g_{\epsilon J|H}}\right)^{-1}
  = \left(1 + \frac{\epsilon}{\g_{J|H}}\right)^{-1}
\,.
\ee
Then, for small $\epsilon$, 
\be
  \g_{H + \epsilon J|H} \geq  1-\frac{\epsilon}{\g_{J|H}} + O(\epsilon^2)
\,.
\ee
Similarly, for simulation (\ref{eq:back}) we arrive at
\be
  \g_{H|H+\epsilon J} \geq  1-\frac{\epsilon}{\g_{-\!J|H+\epsilon J}}
	+ O(\epsilon^2)
\,.
\ee
Furthermore, let $g_0 := \min_{\;0 \leq \epsilon \leq \epsilon_0}
(\g_{\!J|H}^{-1}, \g_{\!-\!J|H+\epsilon J}^{-1})$ for any fixed
$\epsilon_0 > 0$.
Then the Axiom implies 
\bea
  K(H) &\geq&  (1-\epsilon \, g_0)K(H+\epsilon J)  + O(\epsilon^2)\\
  K(H+\epsilon J) &\geq&  (1-\epsilon \, g_0) K(H) + O(\epsilon^2)
\,
\eea
and 
\be
  |K(H)-K(H+\epsilon J)| \leq \epsilon \, g_0 \bar{K} + O(\epsilon^2)
\,,
\ee
where $\bar{K}:=\max_{\; 0 \leq \epsilon \leq \epsilon_0}
K(H\!+\!\epsilon J)$, which proves the continuity of $K$.
\end{proof}

For bipartite tensor product Hamiltonians, the following additional
properties hold.

\begin{prop}[Inverse Hamiltonian]
\label{prop:inve}
For any product Hamiltonian $H=H_A\otimes H_B$,
  \be
    K(-H) = K(H)
  \,.
  \ee
\end{prop}

\begin{prop}[Complex conjugation]
\label{prop:com}
  For any product Hamiltonian $H=H_A\otimes H_B$,
  \be
    K(H^*) = K(H)
  \,.
  \ee
\end{prop}

\begin{prop}[Exchange]
\label{prop:exc}
  For any product Hamiltonian $H=H_A\otimes H_B$ and for $H^{\diamond}$
  the same Hamiltonian acting on exchanged systems,
  \be
    K(H^{\diamond}) = K(H)
  \,.
  \ee
\end{prop}

\begin{prop}[Additivity]
\label{prop:add} 
  For product Hamiltonians $H_1$ and $H_2$ acting on different pairs
  of systems $A_1B_1$ and $A_2B_2$,
\be
    K(H_1\boxplus H_2) = K(H_1) + K(H_2)
  \,.
  \ee
\end{prop}
\noindent
Properties \ref{prop:inve}--\ref{prop:exc} follow from Corollary
\ref{coro:inverse}.  Property \ref{prop:add} follows from Lemma
\ref{lm:three}.

The above Axiom relates the notion of strength measure to that of
Hamiltonian simulation. The following lemma shows that identifying
strength measures is of interest in order to establish bounds on
simulation rates---and thereby also bounds on Hamiltonian capacities,
as discussed in Sec.~\ref{sec:intro}.

\begin{lemma}
The optimal simulation rate $\g_{H'|H}$ corresponds to an optimization
over all possible strength measures $K$, 
\be
  \g_{H'|H} = \min_{K} \frac{K(H)}{K(H')}
\,.
\label{eq:max}
\ee
\end{lemma}

\begin{proof}
From the Axiom, $\g_{H'|H} \leq \frac{K(H)}{K(H')}$ for any strength
measure $K$. In addition, the minimum is achieved by the strength
measure $K(H)=\g_{H'|H}$.
\end{proof}

The many functions that fulfill the Axiom form a convex cone.  In
other words, if the functions $K_1$ and $K_2$ satisfy the Axiom, then
so does $c_1K_1+c_2K_2$ for any $c_1,c_2 \geq 0$.  Of special interest
is the subset of extremal strength measures, i.e., those that cannot
be expressed as a positive sum of others, since we can restrict the
optimization in (\ref{eq:max}) to such functions.  
Ideally, we would like to find a finite subset of extremal strength
measures that form a {\em complete set}, in that optimization over
this set gives the optimal rate $\g_{H'|H}$ for all $H,H'$.  We will
describe two examples of complete sets of extremal strength measures.

The first case corresponds to the simulation of two-qubit Hamiltonians
by two-qubit Hamiltonians.  Any such Hamiltonian can be written, up to
local terms and local unitary transformations, as \cite{DVCLP01}
\be
  H = \sum_{i=x,y,x} \lambda_i \, \sigma_i \otimes \sigma_i \,, \quad
  \lambda_x \geq \lambda_y \geq |\lambda_z|
\,.
\ee
The set of strength measures $K_i$,
\bea
  K_1 &:=& \lambda_x \\
  K_2 &:=& \lambda_x+\lambda_y-\lambda_z \\
  K_3 &:=& \lambda_x+\lambda_y+\lambda_z
\,,
\eea
is a complete set of extremal strength measures in that any optimal rate
$\g_{H'|H}$ can be obtained from the optimization \cite{BCL02} 
\be
  \g_{H'|H} = \min \left\{\frac{K_1(H)}{K_1(H')},
                          \frac{K_2(H)}{K_2(H')},
                          \frac{K_3(H)}{K_3(H')}\right\}
\,.
\ee

The second case corresponds to the simulation of product Hamiltonians
by product Hamiltonians.  We have seen that in this case the
simulation can always be made reversible.  This implies that, up to a
multiplicative constant, there is a unique strength measure for
product Hamiltonians,
\be
  K_{\otimes}(H) := {1 \over 4} \Delta_A\Delta_B
\,.
\ee
Indeed, because of reversibility, any function $K$ fulfilling the Axiom
satisfies
\be
  K(H) = \g_{H'|H} K(H')
\ee
for all product Hamiltonians $H,H'$, so that
\be
  {K(H) \over K(H')} = \g_{H'|H} = {K_{\otimes}(H) \over K_{\otimes}(H')}
\,.
\ee

\section{Discussion}
\label{sec:discussion}

We have seen that all tensor product Hamiltonians can simulate each
other reversibly, so that their nonlocal properties are characterized
entirely by the quantity $K_{\otimes}(H)$ given in (\ref{eq:Ising}).  
This is an example of lossless interconversion of resources, an
appealing situation in information theory.  A related example is the
problem of communication through a one-way classical channel.  By
Shannon's noisy coding theorem \cite{Sha48} together with the reverse
Shannon theorem \cite{BSST01}, all classical channels can simulate
each other reversibly (in the presence of free shared randomness), and
hence they can be characterized entirely in terms of a single
quantity, their capacity.  
Similarly, in the presence of free shared entanglement, all one-way
quantum channels can simulate each other reversibly (at least on
certain input ensembles \cite{BSW}), and thus they are characterized
entirely in terms of their entanglement-assisted capacity for sending
classical information.

In Section \ref{sec:ent}, we saw how Theorem \ref{thm:product} can be
used to extend previous results for two-qubit Hamiltonians to product
Hamiltonians.  Another such extension can be obtained for the problem
of using bipartite Hamiltonians to simulate bipartite unitary gates.
In the case of two-qubit systems, it is known how to optimally produce
any two-qubit gate using any two-qubit Hamiltonian
\cite{VHC02,HVC02,CHN}.  Since all product Hamiltonians are
equivalent to some multiple of the Ising interaction, this result
immediately provides the optimal way to use any product Hamiltonian to
simulate any two-qubit unitary gate, such as the controlled-not gate.

In view of our results, it will be interesting to improve our
understanding of the properties of the Ising interaction.  For
example, as previously mentioned, a calculation of the communication
capacity of the Ising interaction would provide a formula for the
communication capacity of all product Hamiltonians.

It is also interesting to consider Hamiltonian simulation in the
multipartite case.  All of the Rules from Sec.~\ref{sec:sim} have
multipartite analogues, and much of the general discussion on strength
measures for Hamiltonians from Sec.~\ref{sec:strength} can be carried
over as well.  Furthermore, Theorem \ref{thm:product} can be
generalized to more than two parties in the special case in which the
individual tensor factors are traceless.  However, much remains to be
done in the general multipartite case.

Of course, the set of tensor product Hamiltonians is clearly a special
subset of all bipartite Hamiltonians, and thus may not be
representative of the general problem of bipartite Hamiltonian
simulation.  For example, we have seen that product Hamiltonians admit
a total order, whereas even in the two-qubit case, general
Hamiltonians only admit a partial order.  Also, note that for product
Hamiltonians, $H$ and $-H$ are locally equivalent, so that in
particular, $E_H = E_{-H}$.  However, while this is true for all two
qubit Hamiltonians, numerical evidence suggests that it is not true in
general \cite{CLS}.  Understanding optimal Hamiltonian simulation
and the capacities of Hamiltonians in the general case remains an
interesting open problem.

\acknowledgments

AMC received support from the Fannie and John Hertz Foundation, and
was also supported in part by the Cambridge--MIT Foundation, by the
Department of Energy under cooperative research agreement
DE-FC02-94ER40818, and by the National Security Agency and Advanced
Research and Development Activity under Army Research Office contract
DAAD19-01-1-0656.  DWL and GV received support from the US National
Science Foundation under grant number EIA-0086038.


\appendix
\section*{Appendix: Catalytic simulation of a sum of two product
Hamiltonians with the same extremal eigenspace}

As mentioned in Sec.~\ref{sec:equiv}, Theorem \ref{thm:product} can be
extended when catalytic entanglement is available.  Consider a sum of
two tensor product Hamiltonians,
\be
  H = J_{A} \otimes J_{B} + G_{A} \otimes G_{B}
\,.
\ee
Let $J_{\!A}^{\rm \,e}$ denote the restriction of $J_{A}$ to the
subspace corresponding to its two extremal eigenvalues.  By Rule
\ref{rule:local}, $J_{\!A}^{\rm \, e}$ can be assumed to be traceless.
Let $G_{\!A}^{\rm \, e}$, $J_{\!B}^{\rm \, e}$, $G_{\!B}^{\rm \, e}$
be similarly defined.
In terms of these Hamiltonians, we have
\begin{corollary}
\label{cor:xxyy}
  Given the resource of catalytic entanglement, $H = J_A \otimes J_B +
  G_A \otimes G_B$ is locally equivalent to $[(\Delta_J^2 +
  \Delta_G^2)/4] \, H_{\rm Ising}$ if the following conditions hold:
    ($i$)  $J_{\!A}^{\rm \, e}$ and $G_{\!A}^{\rm \, e}$ are
           supported on the same $2$-dimensional Hilbert space, and
           similarly for $J_{\!B}^{\rm e}$ and $G_{\!B}^{\rm \, e}$.
    ($ii$) $\tr J_{\!A}^{\rm \, e} G_{\!A}^{\rm \, e} = 
            \tr J_{\!B}^{\rm \, e} G_{\!B}^{\rm \, e}$.
\end{corollary}

\begin{proof}
$[(\Delta_J^2 + \Delta_G^2)/4] H_{\rm Ising}$ can simulate $H$
term-wise using Lemma \ref{lem:productIsing} and Rule \ref{rule:sum},
with no need for catalytic entanglement.

The following procedure uses $H$ to simulate $[(\Delta_J^2 +
\Delta_G^2)/4] \, H_{\rm Ising}$: 
\begin{enumerate}
\item 
Following Rule \ref{rule:reduction},
Alice and Bob restrict to the extremal eigenspace, which is common to
both terms in $H$ by condition ($i$).  This preserves the extremal
eigenvalues.  The resulting Hamiltonian is essentially a two-qubit
Hamiltonian.  
\item 
We can assume $J_{\!A}^{\rm \, e} \otimes J_{\!B}^{\rm \, e} =
(\Delta_J^2/4) \, \s{zA} \otimes \s{zB}$ by a local change of basis.
This can be chosen so that $G_{\!A}^{\rm \, e} \otimes G_{\!B}^{\rm \,
e} = (\Delta_G^2/4) \, (\cos \theta \, \s{zA} + \sin \theta \, \s{xA})
\otimes (\cos \theta \, \s{zB} + \sin \theta \, \s{xB})$ for some
$\theta$ because of condition ($ii$).
\item 
A further local change of basis takes $J_{\!A}^{\rm \, e} \otimes
J_{\!B}^{\rm \, e} + G_{\!A}^{\rm \, e} \otimes G_{\!B}^{\rm \, e}$ to
its normal form $(\Delta_x^2/4) \, \s{xA} \otimes \s{xB} +
(\Delta_z^2/4) \, \s{zA} \otimes \s{zB}$~\cite{DVCLP01}, where
$\Delta_x^2 + \Delta_z^2 = \Delta_J^2 + \Delta_G^2$.
\item 
Finally, $(\Delta_x^2/4) \, \s{xA} \otimes \s{xB} + (\Delta_z^2/4) \,
\s{zA} \otimes \s{zB}$ can simulate $[(\Delta_x^2 + \Delta_z^2)/4] \,
\s{zA} \otimes \s{zB}$ using catalytic entanglement \cite{VC02a}.
\end{enumerate}
This completes the proof.
\end{proof}

This result allows us to calculate the entanglement capacities of the
relevant Hamiltonians. If $H$ satisfies conditions ($i$) and ($ii$)
of Corollary \ref{cor:xxyy}, then
\be
  E_H = {\alpha \over 4} (\Delta_J^2 + \Delta_G^2)
\,.
\ee
There is an input state that achieves $E_H$ without making
use of ancillas (and in particular, without using catalytic
entanglement), so $\Gamma_H = E_H$.  As mentioned in
Sec.~\ref{sec:ent}, this generalizes the case $H = \mu_x \, \s{x}
\otimes \s{x} + \mu_y \, \s{y} \otimes \s{y}$ considered in
\cite{CLVV03}.


\end{document}